%% file: paper.tex
\documentclass[aps,prl,twocolumn,superscriptaddress,showpacs]{revtex4}
\usepackage{graphicx}

\def\met{$E\!\!\!/_T$}

\begin{document}


\title{Limits on Anomalous Triple Gauge Couplings in $p\bar{p}$
Collisions at $\sqrt{s}=1.96$ TeV}

\input{authors}




\date{\today}

\begin{abstract}
\input{abstract}
\end{abstract}


\maketitle

\input{body}
\bibliography{basename of .bib file}

\input{ref.tex}
\end{document}

%% file: authors.tex
\affiliation{Institute of Physics, Academia Sinica, Taipei, Taiwan 11529, Republic of China} 
\affiliation{Argonne National Laboratory, Argonne, Illinois 60439} 
\affiliation{Institut de Fisica d'Altes Energies, Universitat Autonoma de Barcelona, E-08193, Bellaterra (Barcelona), Spain} 
\affiliation{Baylor University, Waco, Texas  76798} 
\affiliation{Istituto Nazionale di Fisica Nucleare, University of Bologna, I-40127 Bologna, Italy} 
\affiliation{Brandeis University, Waltham, Massachusetts 02254} 
\affiliation{University of California, Davis, Davis, California  95616} 
\affiliation{University of California, Los Angeles, Los Angeles, California  90024} 
\affiliation{University of California, San Diego, La Jolla, California  92093} 
\affiliation{University of California, Santa Barbara, Santa Barbara, California 93106} 
\affiliation{Instituto de Fisica de Cantabria, CSIC-University of Cantabria, 39005 Santander, Spain} 
\affiliation{Carnegie Mellon University, Pittsburgh, PA  15213} 
\affiliation{Enrico Fermi Institute, University of Chicago, Chicago, Illinois 60637} 
\affiliation{Comenius University, 842 48 Bratislava, Slovakia; Institute of Experimental Physics, 040 01 Kosice, Slovakia} 
\affiliation{Joint Institute for Nuclear Research, RU-141980 Dubna, Russia} 
\affiliation{Duke University, Durham, North Carolina  27708} 
\affiliation{Fermi National Accelerator Laboratory, Batavia, Illinois 60510} 
\affiliation{University of Florida, Gainesville, Florida  32611} 
\affiliation{Laboratori Nazionali di Frascati, Istituto Nazionale di Fisica Nucleare, I-00044 Frascati, Italy} 
\affiliation{University of Geneva, CH-1211 Geneva 4, Switzerland} 
\affiliation{Glasgow University, Glasgow G12 8QQ, United Kingdom} 
\affiliation{Harvard University, Cambridge, Massachusetts 02138} 
\affiliation{Division of High Energy Physics, Department of Physics, University of Helsinki and Helsinki Institute of Physics, FIN-00014, Helsinki, Finland} 
\affiliation{University of Illinois, Urbana, Illinois 61801} 
\affiliation{The Johns Hopkins University, Baltimore, Maryland 21218} 
\affiliation{Institut f\"{u}r Experimentelle Kernphysik, Universit\"{a}t Karlsruhe, 76128 Karlsruhe, Germany} 
\affiliation{High Energy Accelerator Research Organization (KEK), Tsukuba, Ibaraki 305, Japan} 
\affiliation{Center for High Energy Physics: Kyungpook National University, Taegu 702-701, Korea; Seoul National University, Seoul 151-742, Korea; SungKyunKwan University, Suwon 440-746, Korea} 
\affiliation{Ernest Orlando Lawrence Berkeley National Laboratory, Berkeley, California 94720} 
\affiliation{University of Liverpool, Liverpool L69 7ZE, United Kingdom} 
\affiliation{University College London, London WC1E 6BT, United Kingdom} 
\affiliation{Centro de Investigaciones Energeticas Medioambientales y Tecnologicas, E-28040 Madrid, Spain} 
\affiliation{Massachusetts Institute of Technology, Cambridge, Massachusetts  02139} 
\affiliation{Institute of Particle Physics: McGill University, Montr\'{e}al, Canada H3A~2T8; and University of Toronto, Toronto, Canada M5S~1A7} 
\affiliation{University of Michigan, Ann Arbor, Michigan 48109} 
\affiliation{Michigan State University, East Lansing, Michigan  48824} 
\affiliation{University of New Mexico, Albuquerque, New Mexico 87131} 
\affiliation{Northwestern University, Evanston, Illinois  60208} 
\affiliation{The Ohio State University, Columbus, Ohio  43210} 
\affiliation{Okayama University, Okayama 700-8530, Japan} 
\affiliation{Osaka City University, Osaka 588, Japan} 
\affiliation{University of Oxford, Oxford OX1 3RH, United Kingdom} 
\affiliation{University of Padova, Istituto Nazionale di Fisica Nucleare, Sezione di Padova-Trento, I-35131 Padova, Italy} 
\affiliation{LPNHE, Universite Pierre et Marie Curie/IN2P3-CNRS, UMR7585, Paris, F-75252 France} 
\affiliation{University of Pennsylvania, Philadelphia, Pennsylvania 19104} 
\affiliation{Istituto Nazionale di Fisica Nucleare Pisa, Universities of Pisa, Siena and Scuola Normale Superiore, I-56127 Pisa, Italy} 
\affiliation{University of Pittsburgh, Pittsburgh, Pennsylvania 15260} 
\affiliation{Purdue University, West Lafayette, Indiana 47907} 
\affiliation{University of Rochester, Rochester, New York 14627} 
\affiliation{The Rockefeller University, New York, New York 10021} 
\affiliation{Istituto Nazionale di Fisica Nucleare, Sezione di Roma 1, University of Rome ``La Sapienza," I-00185 Roma, Italy} 
\affiliation{Rutgers University, Piscataway, New Jersey 08855} 
\affiliation{Texas A\&M University, College Station, Texas 77843} 
\affiliation{Istituto Nazionale di Fisica Nucleare, University of Trieste/\ Udine, Italy} 
\affiliation{University of Tsukuba, Tsukuba, Ibaraki 305, Japan} 
\affiliation{Tufts University, Medford, Massachusetts 02155} 
\affiliation{Waseda University, Tokyo 169, Japan} 
\affiliation{Wayne State University, Detroit, Michigan  48201} 
\affiliation{University of Wisconsin, Madison, Wisconsin 53706} 
\affiliation{Yale University, New Haven, Connecticut 06520} 
\author{T.~Aaltonen}
\affiliation{Division of High Energy Physics, Department of Physics, University of Helsinki and Helsinki Institute of Physics, FIN-00014, Helsinki, Finland}
\author{A.~Abulencia}
\affiliation{University of Illinois, Urbana, Illinois 61801}
\author{J.~Adelman}
\affiliation{Enrico Fermi Institute, University of Chicago, Chicago, Illinois 60637}
\author{T.~Affolder}
\affiliation{University of California, Santa Barbara, Santa Barbara, California 93106}
\author{T.~Akimoto}
\affiliation{University of Tsukuba, Tsukuba, Ibaraki 305, Japan}
\author{M.G.~Albrow}
\affiliation{Fermi National Accelerator Laboratory, Batavia, Illinois 60510}
\author{S.~Amerio}
\affiliation{University of Padova, Istituto Nazionale di Fisica Nucleare, Sezione di Padova-Trento, I-35131 Padova, Italy}
\author{D.~Amidei}
\affiliation{University of Michigan, Ann Arbor, Michigan 48109}
\author{A.~Anastassov}
\affiliation{Rutgers University, Piscataway, New Jersey 08855}
\author{K.~Anikeev}
\affiliation{Fermi National Accelerator Laboratory, Batavia, Illinois 60510}
% ...
\collaboration{CDF Collaboration\footnote{With visitors from $^a$University of Athens, 15784 Athens, Greece, 
$^b$University of Bristol, Bristol BS8 1TL, United Kingdom, 
$^c$University Libre de Bruxelles, B-1050 Brussels, Belgium, 
$^d$Cornell University, Ithaca, NY  14853, 
$^e$University of Cyprus, Nicosia CY-1678, Cyprus, 
$^f$University College Dublin, Dublin 4, Ireland, 
$^g$University of Edinburgh, Edinburgh EH9 3JZ, United Kingdom, 
$^h$University of Heidelberg, D-69120 Heidelberg, Germany, 
$^i$Universidad Iberoamericana, Mexico D.F., Mexico, 
$^j$University of Manchester, Manchester M13 9PL, England, 
$^k$Nagasaki Institute of Applied Science, Nagasaki, Japan, 
$^l$University de Oviedo, E-33007 Oviedo, Spain, 
$^m$University of London, Queen Mary College, London, E1 4NS, England, 
$^n$University of California Santa Cruz, Santa Cruz, CA  95064, 
$^o$Texas Tech University, Lubbock, TX  79409, 
$^p$University of California, Irvine, Irvine, CA  92697, 
$^q$IFIC(CSIC-Universitat de Valencia), 46071 Valencia, Spain, 
}}
\noaffiliation

%% file: abstract.tex
We present a search for anomalous triple gauge couplings (ATGC)
in $WW$ and $WZ$ boson production.
The boson pairs are produced in $p\bar{p}$ collisions at $\sqrt{s}=1.96$ TeV,
and the data sample corresponds to 350 pb$^{-1}$ of
integrated luminosity collected with the CDF II detector at the Fermilab 
Tevatron. In this search one $W$ decays to leptons,
and the other boson ($W$ or $Z$) decays hadronically.
Combining with a previously published CDF measurement of $W\gamma$
boson production yields ATGC limits of $-0.18 < \lambda < 0.17$ and
$-0.46 < \Delta\kappa < 0.39$ at the 95\% confidence level,
using a cut-off scale $\Lambda=1.5$ TeV.

%% file: body.tex
In the standard model (SM), the non-Abelian nature of the electroweak
field theory predicts interactions between the massive gauge bosons.
The resulting triple and quartic boson vertices and couplings are
restricted by electroweak symmetry \cite{ref:mandl}.
By experimentally measuring the strengths of these couplings we can
test the SM and constrain possible deviations from it.
One way to test a more general gauge interaction hypothesis is to
formulate an effective theory and keep only the leading-order (LO)
operators.
This introduces two free parameters $\lambda$ and
$\Delta\kappa=\kappa-1$ under the assumption of equal $WWZ$ and $WW\gamma$
parameters~\cite{Hagiwara:1989mx}, where $\lambda=\Delta\kappa=0$
corresponds to the SM.

At the Fermilab Tevatron $p\bar{p}$ collider the ATGC
contribution to the cross section for diboson production is
a function of the parton center-of-mass energy $\sqrt{\hat{s}}$ and,
in order to maintain unitarity in the model at high energies, a form factor
\begin{eqnarray}
\lambda(\hat{s})=\frac{\lambda}{(1+\hat{s}/\Lambda^2)^2}, \hspace{3mm}
\Delta\kappa(\hat{s})=\frac{\Delta\kappa}{(1+\hat{s}/\Lambda^2)^2},
\end{eqnarray}
is introduced. We set the cut-off energy $\Lambda$ to 1.5 TeV to preserve 
unitarity at the energies reached by the Tevatron \cite{Hagiwara:1989mx}.
The ATGC are enhanced at
large scattering angles \cite{Hagiwara:1989mx}, making observables which are
proportional to the gauge boson transverse momentum particularly effective. 

In this Letter, we present a search for ATGC in $p\bar{p}$
collisions using events in which a $W$ decays to an electron or muon and its
associated neutrino and the other boson ($W$ or $Z$) decays hadronically.
In what follows we refer to this event signature as $\ell\nu jj$. 
The search was performed with the signal region blinded until all selection
criteria and background determinations were fixed.
Since the detector dijet mass resolution does not permit separation of
hadronic decays of $Z$ and $W$ bosons, we combine both $WW$ and $WZ$ in this
analysis. The $\ell\nu jj$ channel has several good features:
by identifying the $p_T$~\cite{ref:coord} of the $\nu$ with $E\!\!\!/_T$,
the $\ell\nu jj$ decay mode allows for a full reconstruction
of the gauge boson transverse momentum; it utilizes the high branching
fraction of $W$ and $Z$ into quarks; the leptonic decay of the
boson gives a clear signature on which to trigger.
However, due to the large $W$+jets background and
the limited dijet mass resolution, no observation has yet been
made of $WW$ or $WZ$ production in the $\ell\nu jj$ decay mode at hadron
colliders. In this analysis we measure the transverse momentum ($p_T^W$)
distribution for $W \rightarrow \ell\nu$. The SM has strong cancellations
between the s-channel and the u- or t-channel diagrams at high $p_T^W$,
and any ATGC will tend to reduce the
cancellations and substantially increase the cross section.

%
%
Previous limits on ATGC in the $\ell\nu jj$ channel in $p\bar{p}$ collisions
at $\sqrt{s}=1.8$ TeV have been reported by the CDF Collaboration, which
obtained 95\% confidence limits (C.L.) of $-1.11<\lambda<1.27$ and
$-0.81<\Delta\kappa<0.84$ \cite{ref:cdf_atgc1}.
The D\O\ collaboration, using a larger data sample,
reported 95\% C.L. limits of  $-0.36<\lambda<0.39$ and
$-0.47<\Delta\kappa<0.63$~\cite{Abbott:1999aj}.
Results on ATGC have also been produced at LEP \cite{Group:2006mx}.
The LEP values are more precise, but the
parameters fitted are not directly comparable as they are obtained at
fixed $\sqrt s$ values up to a maximum of 209 GeV, without the use of form
factors of the type given in Eq. (1). For the Tevatron the spectrum in
 $\sqrt{\hat{s}}$ extends well beyond the reach of LEP, and these studies are
potentially sensitive to the direct production of any new physics beyond
the SM up to the kinematic limit.

The CDF II detector~\cite{ref:CDF} is an approximately  
azimuthally and forward-backward symmetric
apparatus centered on the $p\bar{p}$ interaction region, and consists
of a magnetic spectrometer surrounded by calorimeters and muon chambers.
Charged tracks are detected using a 96-layer
open-cell cylindrical drift chamber (COT) in a $1.4$~${\rm T}$ solenoidal
magnetic field. Isolated high-momentum tracks are reconstructed in the COT
with an efficiency close to $100 \%$ in the pseudorapidity 
region $|{\eta}|<1$.
Electromagnetic and hadronic calorimeters surround the tracking system. 
The calorimeters are segmented into projective geometry towers and cover
the region $|{\eta}|<3.6$. The central and forward 
electromagnetic calorimeters are lead-scintillator
sampling devices, instrumented with proportional (central) and scintillating
strip (forward) detectors that measure the position and transverse profile of
electromagnetic showers at the position of the shower maximum.
The hadron calorimeters are iron-scintillator sampling detectors.
Muon drift chambers surround the calorimeters and, for this analysis, provide
muon identification  for $|{\eta}|<1$.
Gas Cherenkov counters in the region $3.7<|\eta|< 4.7$ measure the average
number of inelastic $p\bar{p}$ collisions per bunch crossing in order to
compute the luminosity to an accuracy of $6\%$~\cite{ref:lumi}.

%
%
The trigger system selects events with a central ($|{\eta}|<1$) electron
candidate with $E_{T}>18$~${\rm GeV}$ or a
muon candidate with $p_{T}>18$~GeV/$c$. Events reconstructed offline
are required to contain an electron candidate with $E_T>25$ GeV or a muon
candidate with $p_T>20$ GeV/$c$, to ensure a trigger
efficiency sufficiently constant in $E_T$ and $p_T$.
The trigger and lepton identification criteria are described in
Ref.~\cite{Acosta:2004hw}. The data presented in this Letter correspond
to a total integrated luminosity of
$350\pm 20$ pb$^{-1}$ for the electron sample and $330\pm 20$ pb$^{-1}$ for
the muon sample.

Candidate events from $W$ leptonic decays are selected by requiring the
$E\!\!\!/_T$, corrected for muons and jets, 
to be greater than $25$~${\rm GeV}$. 
To further reduce the QCD multijet background, the $W$
transverse mass is required to satisfy 
$m_T^W=\sqrt{2E_T E\!\!\!/_T(1-\cos\Delta\phi)}> 25$ GeV/$c^2$, where
$\Delta\phi$ is the angle between the lepton candidate momentum vector
and the $E\!\!\!/_T$ vector in the transverse plane.
Events are required to have only one
lepton candidate, no identified cosmic ray muon, and no identified photon
conversion electron. Most $Z\to~l^{+}l^{-}$ events that remain in the sample
are rejected with a veto algorithm designed to identify event topologies
consistent with a partially reconstructed second lepton \cite{Acosta:2004hw}.

%
%
The $\ell\nu jj$ event selection proceeds by requiring two or more jets.
Jets are reconstructed using an iterative seed-based cone algorithm
\cite{Abe:1991ui} that
clusters energies measured in individual calorimeter
towers. The jets are defined by the cone algorithm parameter
$\Delta R=\sqrt{\Delta\phi^2+\Delta\eta^2} \leq 0.4$. The measured
jet $E_{T}$ is corrected for
calorimeter response and energy contributions from
additional $p\bar{p}$ interactions in the same bunch
crossing~\cite{Bhatti:2005ai}. For this analysis, jets must have corrected
$E_T>20$~${\rm GeV}$ and $|\eta|<2$. To ensure that the event
kinematics are well measured, events are rejected if any jet lies
within $\Delta R=0.5$ of the $W$ decay lepton or another jet.

%
%
After event selection, 929 events in the electron channel and
688 events in the muon channel remain within the signal region. The signal
region is defined as 56~$< M_{jj} <$~112 GeV/$c^2$, where
$M_{jj}$ is the
invariant mass of the two leading jets. Table \ref{tab:ev} shows the
expected and observed numbers of events, where the expected number of events
takes account of kinematic and geometric acceptance, lepton identification
efficiencies, and trigger efficiencies.
%
%
The expected numbers of $WW$ and $WZ$ events are
calculated using information from both data and
Monte Carlo simulations. For the SM diboson production signals we 
use the {\sc pythia} Monte Carlo
generator \cite{ref:pythia} and {\sc geant}-based detector 
simulation \cite{ref:geant}
to predict the accepted event yields.
The lepton efficiencies from these simulations are scaled to match the values
measured from $Z\rightarrow l^+l^-$ events.
The central muon efficiencies require a scale factor correction of
0.874$\pm$0.009. All other systems have scale factors close to unity.
The expected event yields are based on the
total signal cross sections  of 12.4$\pm$0.8 pb ($WW$)
and 4.0$\pm$0.3 pb ($WZ$)
from next-to-leading order (NLO) predictions \cite{ref:xsww}.
Sources of systematic uncertainty on the expected number of events 
include jet $E_T$ corrections (11\%), estimated from data events with two jets,
$\gamma+$jet events, and Monte Carlo simulations tuned to
data~\cite{Bhatti:2005ai};
higher-order QCD radiation estimated from variations of the initial
and final state showering model (10\%); NLO cross section normalization, taken
from theory (7\%); luminosity measurements (6\%); and parton distribution
functions (3\%).

%
%
Backgrounds to $WW$ and $WZ$ production in the $\ell\nu jj$
event signature can be classified into three categories:
electroweak (EWK), QCD multijets, and $W+$jets. 
EWK backgrounds contain real leptons from $W$ and $Z$ decays and 
include $W\to\tau\nu$, $Z\to l^+l^-$, top pair-production and
single top production. The QCD background arises from multijet
events in which one jet is falsely
identified as a lepton or contains a lepton not from
$W$ or $Z$ decays, and which have mismeasured energy resulting in
large \met. The $W+$jets background corresponds to production of single
$W$ bosons decaying to $e\nu$ or $\mu\nu$ accompanied by
additional jets.

%
%
The EWK backgrounds are estimated from Monte Carlo simulations normalized
to NLO predictions \cite{ref:stop,ref:top,ref:mcfm}, using the same methods
as for $WW$ and $WZ$ signal expectations.

%
%
The QCD background is estimated from data.
Assuming no correlation between event $E\!\!\!/_T$ and
energy in the vicinity of the lepton (lepton isolation energy), we
extrapolate from sideband regions to predict the QCD content
in the signal region, correcting for the EWK
contributions \cite{Acosta:2004hw}.
By using several different ranges of lepton isolation energy and
$E\!\!\!/_T$ we estimate a 40\% QCD normalization uncertainty.

\begin{figure}
\includegraphics[height=.24\textheight]{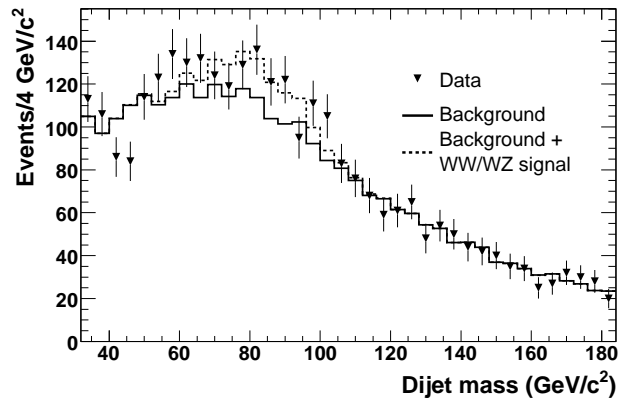}
\caption{
Invariant mass reconstructed from the two jets with highest $E_T$.
The measured data are fitted to the hypothesis of combined signal and
background shapes.\label{fig:wwmass}}
\end{figure}

%
%
The $W+$jets background is simulated using the {\sc alpgen} Monte Carlo 
generator \cite{ref:alpgen},
followed by {\sc herwig} \cite{ref:herwig} for the parton shower and
fragmentation, and full {\sc geant} detector simulation. The
hard scattering process includes two partons in the final state.
To further minimize the systematic
uncertainties, we float the absolute $W+$jets normalization
in a fit to data (Fig. \ref{fig:wwmass}). In the fit, we allow for a linear
dependence of the normalization on the dijet invariant mass ($M_{jj}$),
derived from studies of the effect of renormalization scale variation.
By normalizing $W$+jets to the observed data, we achieve
a normalization uncertainty
of 5\%. This should be contrasted with the 20\% uncertainty on the NLO
cross section that would otherwise have been used to normalize this
contribution.

Before probing the existence of ATGC, we test our sensitivity to SM $WW$ and
$WZ$ production by fitting the dijet mass distribution to the
background plus signal hypothesis in the extended dijet mass region
$32 < M_{jj} < 184$ GeV/$c^2$, using a lowered jet $E_T$ threshold of 15 GeV.
The result of the fit is $109 \pm 110~(stat)
\pm 54~(syst)$ $WW + WZ$ events, consistent with both the SM expectation of
160 events and also with no $WW + WZ$ production. We set a 95\%
C.L. upper limit of 36 pb on the combined $WW + WZ$ production cross section.

%
%
\begin{table}
\caption{Expected and observed data events for the $\ell\nu jj$ $WW$
and $WZ$ search in the signal region. The signal region is defined
as 56 $<  M_{jj} <$ 112 GeV/$c^2$.
$e$+$\mu$ is a correlated
estimation of both channels. The listed uncertainties are
statistical and systematic combined.\label{tab:ev}}
\begin{tabular}{cr@{ }lr@{ }lr@{ }l}
 & \multicolumn{2}{c}{$e$} &  \multicolumn{2}{c}{$\mu$} &  \multicolumn{2}{c}{$e+\mu$} \\
\hline
$WW$ & 44.7 $\pm$ & \phantom{0}7.6 & 34.4 $\pm$ & \phantom{0}5.8 & 79.0 $\pm$ & 13.4 \\
$WZ$ & 6.7 $\pm$ & \phantom{0}1.1 & 5.1 $\pm$ & \phantom{0}0.9 & 11.8 $\pm$ & \phantom{0}2.0 \\
\hline
Signal & 51.4 $\pm$ & \phantom{0}8.7 & 39.5 $\pm$ & \phantom{0}6.7 & 90.9 $\pm$ & 15.5 \\
\hline
$W+$jets & 690.0 $\pm$ & 52.0 & 552.2 $\pm$ & 44.3 & 1242.2 $\pm$ & 65.4 \\
QCD Multijets & 53.7 $\pm$ & 21.5 & 11.9 $\pm$ & \phantom{0}4.8 & 65.6 $\pm$ & 26.2 \\
$t\bar{t}$ & 30.9 $\pm$ & \phantom{0}7.7 & 22.4 $\pm$ & \phantom{0}5.6 & 53.3 $\pm$ & 13.3 \\
$Z+$jets & 16.8 $\pm$ & \phantom{0}3.4 & 26.9 $\pm$ & \phantom{0}5.4 & 43.7 $\pm$ & \phantom{0}8.7 \\
$W(\tau\nu)+$jets & 17.7 $\pm$ & \phantom{0}3.5 & 21.9 $\pm$ & \phantom{0}4.4 & 39.6 $\pm$ & \phantom{0}7.9 \\
Single & 5.1 $\pm$ & \phantom{0}1.3 & 3.6 $\pm$ & \phantom{0}0.9 & 8.7 $\pm$ & \phantom{0}2.2 \\
\hline
Background & 814.1 $\pm$ & 57.0 & 639.0 $\pm$ & 45.5 & 1453.1 $\pm$ & 72.7 \\
\hline
Expected & 865.5 $\pm$ & 57.7 & 678.5 $\pm$ & 46.0 & 1544.0 $\pm$ & 74.3 \\
Observed & 929 & & 688 & & 1617 &\\
\end{tabular}
\end{table}

\begin{figure}
\includegraphics[height=.24\textheight]{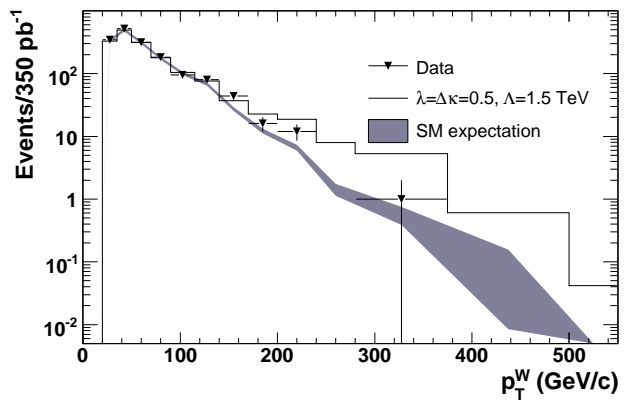}
\caption{$p_T^W$ distribution used for setting the ATGC limits.
Shown are the SM expectation, data points, and one ATGC scenario with
$\lambda=\Delta\kappa=0.5$ and  $\Lambda=1.5$ TeV.
\label{fig:enupt}}
\end{figure}

\begin{figure}
\includegraphics[height=.24\textheight]{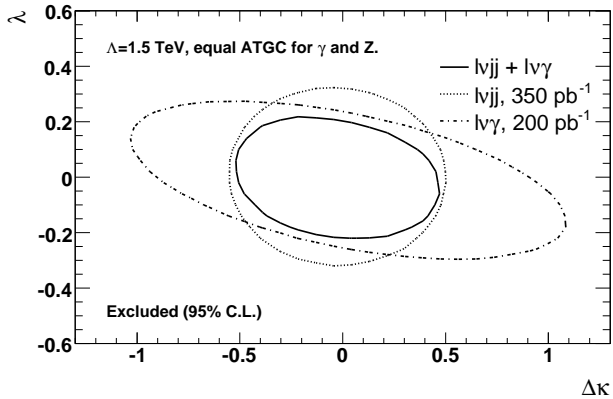}
\caption{Limits at 95\% C.L. on ATGC using $WW$ and $WZ$ decays to
$\ell\nu jj$, $W\gamma$ decays to $\ell\nu\gamma$, and the combination of both
channels.
\label{fig:lvjj}}
\end{figure}

%
%
While we observe no significant evidence of SM $WW + WZ$
production, we can probe ATGC with the $p_T^W$
distribution, for which we expect maximum sensitivity to ATGC. In particular,
ATGC would result in an enhanced cross section at
high $p_T^W$, where
SM backgrounds are small. We perform a binned maximum likelihood fit to the
measured $p_T^W$ spectrum (Fig. \ref{fig:enupt}), and observe no significant
deviation from the SM.
The ATGC signal is simulated at LO using the {\sc mcfm} Monte
Carlo generator \cite{ref:xsww} for the hard scatter process and
fragmented with {\sc pythia}, followed by full {\sc geant} detector simulation.
A grid of points is simulated in the $\lambda$-$\Delta\kappa$ plane, and
the expected numbers of events at each point are fitted to a quadratic form.
We determine a two-dimensional 95\% C.L. interval in 
$\lambda$ and $\Delta\kappa$ corresponding to a change of 3.0 units 
in the logarithm of a binned likelihood combining all 
channels, relative to the maximum of the likelihood in
$\lambda$ and $\Delta\kappa$.
Uncertainties from jet energy scale, renormalization scale,
higher order QCD radiation, parton distribution functions, and luminosity
are included by a convolution of the likelihood with Gaussian-distributed
systematic uncertainties. The dominant systematic uncertainty comes from
the jet energy scale, whose correlations among bins, across channels, and
between signal and background, are taken into account during the convolution
process. The two-dimensional limits are shown
in Fig. \ref{fig:lvjj}. We also present the one-dimensional limits
in Table \ref{tab:atgc}, where one of the anomalous couplings is held fixed
to the SM value.
%
%
\begin{table}
\caption{Allowed ATGC ranges for $\Lambda = 1.5$ TeV at
95\% C.L., fixing the other coupling to the SM value.}
\label{tab:atgc}
\begin{tabular}{ccccc}
 & $\lambda$ & $\Delta\kappa$ \\
\hline
$\ell\nu jj$ & (-0.28, 0.28) & (-0.50, 0.43)  \\
$\ell\nu\gamma$ \cite{ref:wgamma} & (-0.21, 0.19) & (-0.74, 0.73) \\
Combined & (-0.18, 0.17) & (-0.46, 0.39) \\
\end{tabular}
\end{table}

%
%
To increase our sensitivity to ATGC, we combine the $\ell\nu jj$ channel
with a previously published CDF measurement of $W\gamma$ production in the
$\ell\nu\gamma$ channel \cite{ref:wgamma}. The
$\ell\nu\gamma$ data set corresponds to about 200 pb$^{-1}$ of
accumulated data. We use $E_T$ of the $\gamma$ to set limits using the same
procedure as outlined above for the $\ell\nu jj$ channel, with the exception
that the Baur Monte Carlo generator \cite{ref:baur} is used for the ATGC
signal.
Systematic uncertainties for $\ell\nu\gamma$ that enter into the
combination are
signal acceptance (electrons 4.6\%, muons 5\%), signal and background
NLO normalization (7\%), luminosity (6\%), and the rate of
mismeasured photons due to jets. The rate of jets misreconstructed as
photons is estimated from data and binned in $E_T$.
The uncertainties on the electron and
the muon channel acceptances are assumed to be correlated
since they are both dominated by the photon identification uncertainty.
The $\ell\nu\gamma$ and $\ell\nu jj$ channels
are assumed to be uncorrelated except for the luminosity measurements.
The derived $\ell\nu\gamma$ and combined limits are shown in
Table \ref{tab:atgc} and  Fig. \ref{fig:lvjj}.
It can be seen that the $\ell\nu jj$ and $\ell\nu\gamma$ data have
complementary sensitivity to $\Delta\kappa$ and $\lambda$.

%
%
In summary, we set limits on ATGC
in 350 pb$^{-1}$ of $p\bar{p}$ collisions at $\sqrt{s}=1.96$ TeV using
the $p_T^W$ from $\ell\nu jj$ $WW$ and $WZ$ decays. The couplings are
restricted to $-0.28<\lambda<0.28$ and $-0.50<\Delta\kappa<0.43$, at 95\% C.L.
This restriction assumes a
form factor cut-off $\Lambda=1.5$ TeV and equal $Z$ and $\gamma$ couplings.
The limits obtained in the $\ell\nu jj$ channel are improved by
combining with previously published
$W\gamma$ results, yielding the combined limits $-0.18 < \lambda < 0.17$ and
$-0.46 < \Delta\kappa < 0.39$.

We thank the Fermilab staff and the technical staffs of the participating
institutions for their vital contributions. This work was supported by the
U.S. Department of Energy and National Science Foundation; the Italian
Istituto Nazionale di Fisica Nucleare; the Ministry of Education, Culture,
Sports, Science and Technology of Japan; the Natural Sciences and
Engineering Research Council of Canada; the National Science Council of the
Republic of China; the Swiss National Science Foundation; the A.P. Sloan
Foundation; the Bundesministerium f\"ur Bildung und Forschung, Germany; the
Korean Science and Engineering Foundation and the Korean Research Foundation;
the Particle Physics and Astronomy Research Council and the Royal Society, UK;
the Institut National de Physique Nucleaire et Physique des Particules/CNRS;
the Russian Foundation for Basic Research; the Comisi\'on Interministerial
de Ciencia y Tecnolog\'{\i}a, Spain; the European Community's Human Potential
Programme; the Slovak R\&D Agency; and the Academy of Finland. 

%% file: paper.bbl
\begin{thebibliography}{99}
\bibitem{ref:coord}{We use a coordinate system where $\theta$ is the polar
angle to the proton beam, $\phi$ is the azimuthal angle about the beam axis,
and $\eta$ is the pseudorapidity defined as $-\ln(\tan(\theta/2))$. The
transverse momentum of a particle is denoted by $p_{T}\equiv p\sin\theta$.
The analogous quantity defined using energies, $E_{T}\equiv E\sin\theta$,
is called transverse energy. Missing transverse energy, \met, is defined
as the magnitude of $-\sum_{i}E_{T}^{i}\hat{n}_{i}$, where $\hat{n}_{i}$ is a
unit vector in the azimuthal plane that points from the collision vertex to
the $i^{th}$ calorimeter tower.}
\end{thebibliography}
